\documentclass[twocolumn,showpacs,preprintnumbers,amsmath,amssymb]{revtex4}

\usepackage{graphicx}
\usepackage{dcolumn}
\usepackage{bm}

\begin{document}

\preprint{XAS \& XES of doped diamond}

\title{Electronic structures of B-2$p$ and C-2$p$ of boron-doped diamond film \\
by soft X-ray absorption and emission spectroscopy}

\author{Jin Nakamura\email{jin@pc.uec.ac.jp}, Eiki Kabasawa, Nobuyoshi Yamada}%
\affiliation{%
Department of Applied Physics \& Chemistry, The University of Electro-Communications, 
Chofu-shi, Tokyo 182-8585, Japan
}%
\author{Yasuaki Einaga}
\affiliation{
Department Chemistry, Keio University, 
Hiyoshi, Kanagawa 305-8568
}%
\author{Daisuke Saito, Hideo Isshiki, Shigemi Yugo}
\affiliation{
Department of Electreo-Engeering, The University of Electro-Communications, 
Chofu-shi, Tokyo 182-8585, Japan
}%
\author{Rupert C.C. Perera}
\affiliation{
Center for X-ray Optics, Lawrence Berkeley National Laboratory, Berkeley, CA 94720
}%
\date{submitted to Phys. Rev. B, June 14 2004}

\begin{abstract}
X-ray absorption (XAS) and emission (XES) spectroscopy near B-K and C-K edges have been 
performed on metallic ($\sim$1at\%B, B-diamond) and semiconducting ($\sim$0.1at\%B and N, 
BN-diamond) doped-diamond films.  
Both B-K XAS and XES spectra shows metallic partial density of state (PDOS) with the 
Fermi energy of 185.3 eV, and there is no apparent boron-concentration dependence 
in contrast to the different electric property.  
In C-K XAS spectrum of B-diamond, the impurity state ascribed to boron is clearly observed 
near the Fermi level.  
The Fermi energy is found to be almost same with the top of the valence band of non-doped 
diamond, $E_{\rm V}$, 283.9 eV.  
C-K XAS of BN-diamond shows both the B-induced shallow level and N-induced deep-and-broad 
levels as the in-gap states, in which the shallow level is in good agreement with the activation 
energy ($E_{\rm a}$=0.37 eV) estimated from the temperature dependence of the conductivity, 
namely the change in C-2$p$ PDOS of impurity-induced metallization is directly observed.  
The electric property of this diamond is mainly ascribed to the electronic structure of 
C-2$p$ near the Fermi level.  
The observed XES spectra are compared with the DVX$\alpha$ cluster calculation.  
The DVX$\alpha$ result supports the strong  hybridization between B-2$p$ and C-2$p$ observed in 
XAS and XES spectra, and suggests that the small amount of borons ($\leq$1at\%) 
in diamond occupy the substitutional site rather than interstitial site. 
\end{abstract}
\maketitle

\section{\label{intro}Introduction}
Diamond is a very attractive material with industrial applications 
because of its maximum hardness, high surface stability 
(chemical inertness), large energy gap ($\sim$5.5 eV), high thermal conductivity, and so on.  
Boron-doped diamond expands its possibility into an application of electric devices.\cite{gildenblat}  
Lightly boron-doped diamond shows $p$-type character with an activation energy of about 
0.37 eV,\cite{glesener}  and heavily doped diamond shows metallic conductivity.\cite{shimomi}  
Furthermore the recent discovery of the superconductivity of more-heavily boron 
doped diamond brought a new attention to the  problem in the superconductivity of 
impurity-induced metallization in semiconductor,\cite{ekimov}    
However the crystalinity of these heavily doped compounds is not clear 
in contrast to the low ($\leq $ 0.5 \%) doped diamonds.  
It seems that boron atoms occupy the interstitial sites when heavily doped case 
($\sim$4\%)\cite{chen,thonke} and substitute for carbon at low doped case ($\leq $ 0.5 \%).\cite{werner}  
It should be important to clarify the memorable electronic structure of more-heavily doped diamonds, 
but at present the priority study should be that of the low doped diamond due to their crystalinities.    
Therefore, in this paper, we study the electronic structures of low doped diamond with  
the metallic ($\sim$1at\%) and semiconducting ($\sim$0.1at\%) characters.  
The partial density of states (PDOS's) of boron- and carbon-2$p$ using X-ray absorption (XAS) 
and  X-ray emission (XES) spectroscopy near the B-K and C-K edges of these doped diamonds are 
reported.     
XAS and XES near B-K and C-K edges are powerful techniques for direct measurement of  PDOS's 
of dopant-boron and host-carbon, especially for the semiconducting or insulating materials, 
in comparison with the electron spectroscopy.
\section{\label{exp}Experimental}
Highly boron-doped diamond thin films were deposited on Si (100) wafers in a
microwave plasma-assisted chemical vapor deposition (MPCVD) system (ASTeX Corp.).  
Details of the preparation were described elsewhere.\cite{yano}   
A mixture of acetone and methanol in the volume ratio of 9/1 was used as the carbon
source.  
B$_2$O$_3$, the boron source, was dissolved in the acetone-methanol solution 
at a B/C atomic ratio of 1:100.  
This 1 at\% boron-doped diamond, B-diamond, shows metallic conductivity at the room temperature.  
The lightly doped diamond film is synthesized using MPCVD method with a $h$-BN target 
(BN-diamond).\cite{saito}  
The boron and nitrogen concentrations are estimated to be 0.1at\% for both boron and nitrogen 
by SIMS measurements.  
And the electric property is semiconducting with the activation energy $E_a$ of about 0.37 eV.\cite{saito}  
However the value of $E_a$ depends on the impurity concentration, this value is consistent with 
the previous reports of lightly boron-doped diamond film.\cite{gildenblat,werner}  
Several values of the nitrogen-impurity levels (deep $n$-type) are reported.\cite{iakoubovskii}   

Soft X-ray absorption (XAS) and XES measurements were performed at BL-8.0.1\cite{alsbl8} 
of the Advanced Light Source (ALS) in Lawrence Berkeley National Laboratory (LBNL).  
The energy resolutions of the incoming and outgoing X-rays were 0.2$\sim$0.3 eV.  
For the calibrations of the monochromator and spectrometer, $h$-BN, B$_2$O$_3$, HOPG and 
natural diamond were used as the standard samples.\cite{ma,skytt,muramatsu}  
Although all the samples are polycrystals, there is a possibility of orientation.  
In order to check the orientation and the surface $\pi$-resonant state reported in some borides, 
polarization (angle) dependencies of XAS and XES were measured.  
There are no essential differences in XAS and XES spectra among those with different angles, 
which suggests that  there is no orientation and any borides showing the surface $\pi$-resonant state 
in these samples. 
\section{\label{R_D}Results and Discussions}
\subsection{\label{B-K}B-K XAS and XES of doped diamond film}
\begin{figure}
\includegraphics[width=7cm]{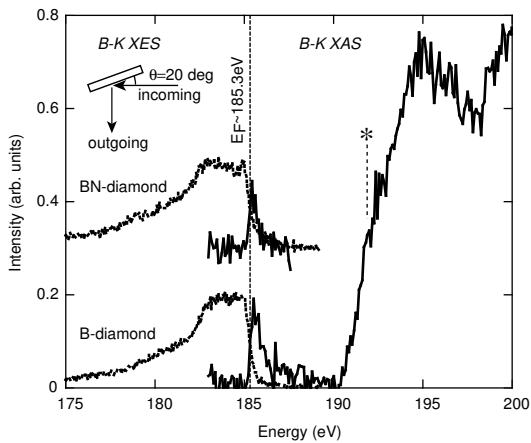}
\caption{\label{Fig1}
B-K XAS and XES spectra of B- and BN-diamonds.  
The incident angle was set to $\theta$=20$^{\circ}$.  
The excitation energy of XES measurement was 200 eV. }
\end{figure}

Figure~\ref{Fig1} shows B-K XAS and XES spectra of B- and BN-diamonds.   
In both compounds, clear metallic states of B-2$p$ are observed, in which both the Fermi 
level are the same with each other to be of 185.3 eV measured from B-1s core level.  
There is a pseudo-gap state between 187 and 190 eV, and the intensity steeply increases 
with an increase of energy with the threshold of 190.5 eV.  
It is noticed that there is no sharp peak at 192 eV (* in Fig.~\ref{Fig1}) which corresponds 
to the surface $\pi$-resonant state of some borides, $h$-BN and B$_2$O$_3$.  
This means there is no trace of these borides in these diamond samples.  

\begin{figure}
\includegraphics[width=7cm]{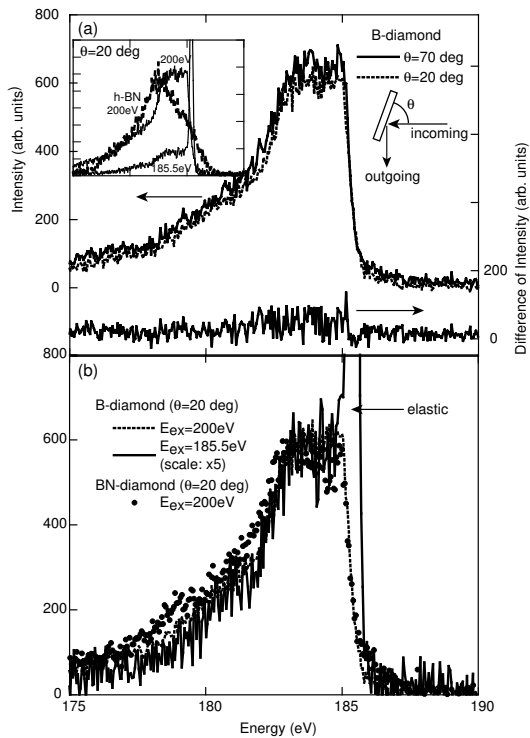}
\caption{\label{Fig2}
B-K XES spectra of B- and BN-diamonds.  
(a) The incident-angle dependence of B-diamond.  The inset shows excitation-energy dependence 
and shows B-K XES spectrum of $h$-BN.  
(b) The excitation-energy dependence of B-diamond, in which the spectrum with $E_{\rm ex}$ of 185.5 eV 
being magnified 5 times.  The spectrum of BN-diamond with $E_{\rm ex}$ of 200 eV is also shown. }
\end{figure}

Figure~\ref{Fig2} shows angular and energy dependence of B-K XES spectra of B- and BN-diamonds.  
Figure~\ref{Fig2}(a) shows two B-K XES spectra of B-diamond with the different incident 
angle $\theta$ of 70$^{\circ}$ and 20$^{\circ}$.  
The excitation energy, $E_{\rm ex}$, is 200 eV.  
These two spectra coincide each other, which indicates that there is no orientation 
of this sample.  
The most important point is the observation of the clear Fermi edge in both samples 
with the same threshold at 185.0 eV in consistent with B-K XAS spectra.  .  
The inset viewgraph shows B-K XES spectrum with $E_{\rm ex}$ of 185.5 eV which 
corresponds to the sharp state near the Fermi level in B-K XAS spectrum (Fig.~\ref{Fig1}).    
In the inset, the spectra of $h$-BN with $E_{\rm ex}$ of 200 eV is also plotted.  
The bonding in $h$-BN is ideal sp$^2$ between B and N having a peak at around 182$\sim$183 eV.  
The observed B-K XES spectra of B-diamond differ from that of $h$-BN, 
but it is hard to say that this spectra is due to sp$^3$ of B in diamond from these results only.  
Figure~\ref{Fig2}(b) shows detailed excitation-energy dependence of B-diamond.  
For the spectrum with $E_{\rm ex}$ of 185.5 eV, the elastic (intense) peak at is 
observed.  
Then we magnified the spectrum with $E_{\rm ex}$ of 185.5 eV 5 times in Fig.~\ref{Fig2}(b).  
The detailed features of these spectra in the energy region of $E \leq $ 184 eV agrees well with 
each other.  
This means that all B-2$p$ has an unique electronic structure.  
Furthermore the spectrum of BN-diamond shows almost same form, which means 
there is no B concentration dependence in these doping region.  
\subsection{\label{C-K_XAS}C-K XAS of doped diamond film}
\begin{figure}
\includegraphics[width=7cm]{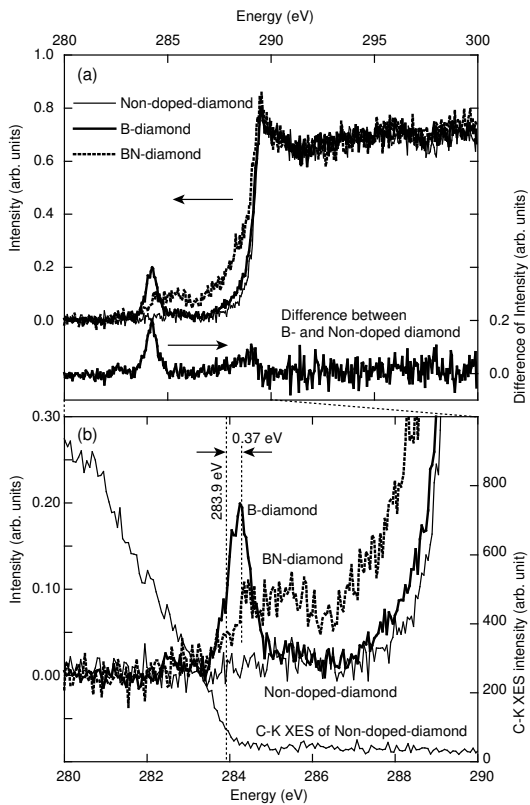}
\caption{\label{Fig3}
C-K XAS spectra of B-, BN- and Non-doped-diamonds.  
(a) overall features of those spectra, (b) detailed spectra of the in-gap states. 
The XAS and XES of non-doped diamond are also shown.  
The dashed lines indicate the top of the valence band of non-doped diamond 
and the impurity level excepted from the activation energy of BN-diamond, respectively. }
\end{figure}

Figure~\ref{Fig3} shows the C-K XAS spectra of B-, BN- and non-doped-diamonds.    
Figure~\ref{Fig3}(a) shows the overall features of C-K XAS spectra.  
The spectrum of non-doped-diamond (thin solid line) shows clear gap state with $E \leq$ 289.1 eV 
which corresponds to the bottom of the conduction band (C.B.).  
On the other hand, the spectra of B- and BN-diamonds show in-gap states.  
For B-diamond (thick solid line), only one peak at 284 eV is observed as the in-gap state.  
The threshold energy of this peak is estimated to about 283.9 eV which is consistent with the 
energy expected from both the observed bottom of C.B. (289.1 eV) and the band gap of 
non-doped diamond.  
In Fig.~\ref{Fig3}(a), the difference between B- and non-doped diamonds is also shown.   
It is clearly seen that the 1-at\%B in diamond makes a metallic state in C-2$p$ PDOS.  
But for BN-diamond (dotted line), the threshold energy shifts higher a little.  
Figure ~\ref{Fig3}(b) shows detailed in-gap states and shows the edge of C-K XES 
of non-doped diamond.  
The threshold energy of the impurity state in XAS of B-diamond, 283.9 eV, is agreement 
with the edge energy in C-K XES of non-doped diamond (the top of the valence band, $E_{\rm v}$).  
It is noted that all the profiles of normal C-K XES of B-, BN- and non-doped diamond 
are almost same (see section~\ref{C-K_XES}).  
In contrast to the metallic B-diamond, BN-diamond seems to has a small gap.  
For this semiconducting sample the activation energy, $E_{\rm a}$, of 0.37 eV was measured from 
the temperature dependence of the conductivity.\cite{saito}  
Then, the shift of 0.3$\sim$0.4 eV between metallic and semiconducting samples is consistently explained 
by the small gap with $E_{\rm a}$ of 0.37 eV.   
In other words, 0.1-at\%B makes a shallow level near the valence band with $E_{\rm a}$ of 0.37 eV.   
In addition, a broad in-gap state spreads over the gap is observed in BN-diamond.  
It is reported that nitrogen-dopant makes several deep levels in diamond\cite{iakoubovskii} and 
the present spectra are similar to the reported C-K XAS spectra for graphite-carbon nitride system.\cite{n-ingap}  
Then this broad in-gap state is ascribed to N-dopant.  
\subsection{\label{C-K_XES}C-K XES of doped diamond film}
\begin{figure}
\includegraphics[width=7cm]{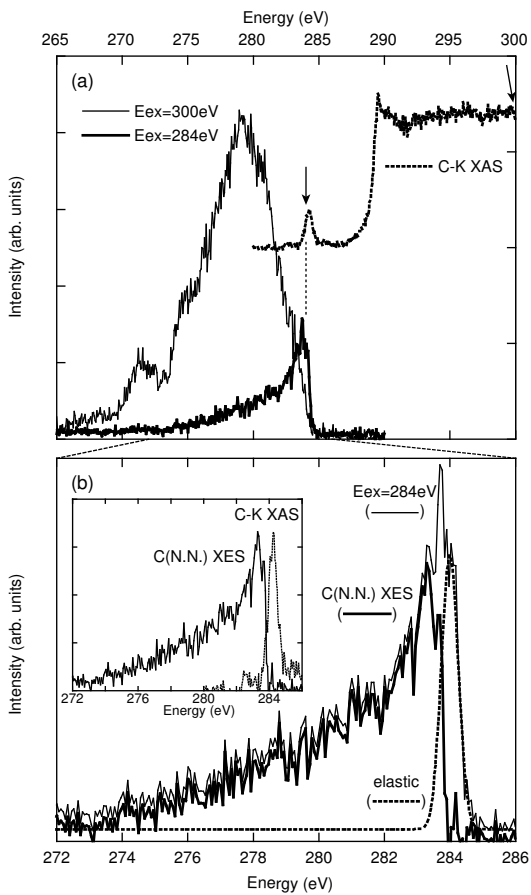}
\caption{\label{Fig4}
C-K XES spectra of B-diamond: 
(a) XES spectra with $E_{\rm ex}$ of 284 eV (near the Fermi energy) and 
of 300 eV.  C-K XAS spectrum of B-diamond is also shown (two arrows indicate 
two $E_{\rm ex}$ positions).  
(b) C(N.N.) XES spectra derived from the subtraction of the elastic peak from 
observed XES spectra with $E_{\rm ex}$ of 284 eV (near the Fermi energy).  
The inset shows C(N.N.) XES and C-K XAS spectra.  
}
\end{figure}

Figure~\ref{Fig4}(a) shows the C-K XES spectra with $E_{\rm ex}$ of 284 eV and 300 eV.  
The spectrum with $E_{\rm ex}$ of 300 eV (dotted line) is almost same as 
the spectrum of the non-doped-diamond.\cite{ma}  
This is consistent with the band calculation results.\cite{oguchi}  
This agreement between doped and non-doped diamond suggests that the host C-1s core level 
does not change by B-doping within the experimental error.  
However, the spectrum with $E_{\rm ex}$ of 284 eV (thick solid line) shows 
a sharp peak at about 284 eV (elastic peak) and a broad tail toward the low energy side.   
Figure~\ref{Fig4}(b) shows the subtraction of the elastic peak from the spectrum.  
The elastic peak was assumed to be gaussian with FWHM of 0.6 eV.\cite{slit}  
The result of the subtraction of the elastic peak is essentially same as before the subtraction 
because the width of elastic peak is narrow.  
Because the unoccupied state, a peak at 284 eV is observed  only in B-doped-diamond, 
the XES spectrum with $E_{\rm ex}$ of 284 eV is a represent of the PDOS of the carbon 
atom hybridized with dopant B, i.e., nearest neighboring (N.N.) carbon from boron.  
Because it is difficult to study the electronic structures of dopant (dilute) boron and the 
neighboring carbon atoms using band calculation, calculations were performed by,  
DVX$\alpha$ method.\cite{dvxa}, a cluster calculation method.   
\subsection{\label{dvxa}DVX$\alpha$ cluster calculation}
The program SCAT\cite{dvxa} was used for DVX$\alpha$ calculation.  
Because the samples are covalent, Madelung potential was not taken into account.  
It is also known that an average of PDOS's of a few atoms near the center of 
the large cluster reproduces well experimental results. 
In the present work, we calculated the PDOS's for non-doped-diamond, and 
two doped diamond cases where boron occupies the interstitial site and substitute's 
carbon site.  
The typical cluster size is about 200 atoms, a limitation due to 
the memory size of the program.  
First, PDOS of non-doped cluster model, C$_{184}$, was calculated and 
the results were compared with the experimental data and the band calculations.  
Although the Fermi levels in both the calculations and observation are not agreed exactly, 
the overall feature of the PDOS's are in agreement with each others.  
For the B doped diamond case, typical cluster models of C$_{174}$BH$_{16}$ and 
C$_{184}$BH$_{12}$ were used for substitutional and interstitial cases, respectively.  
In these cases, the non-doped models, C$_{175}$H$_{16}$ and C$_{184}$H$_{12}$ 
were also applied and the results were confirmed to be same with that of 
C$_{184}$.  
In both B-doped cluster models, boron atom is always set at the center of the clusters.  
In these large clusters, an effect of H-termination is found to be negligible for the PDOS's of 
inner C or B atoms. 
It is noted that a few in-gap states are appeared even in the non-doped case.  
It might be ascribed to the surface states.  
Then in the present DVX$\alpha$ results, the origin of the energy is set to the 
maximum energy of the electron in the occupied states except the meaningless 
surface states, i.e., the energy is measured from the top of the valence band (V.B.), $E_{\rm V}$.  

\begin{figure}
\includegraphics[width=7cm]{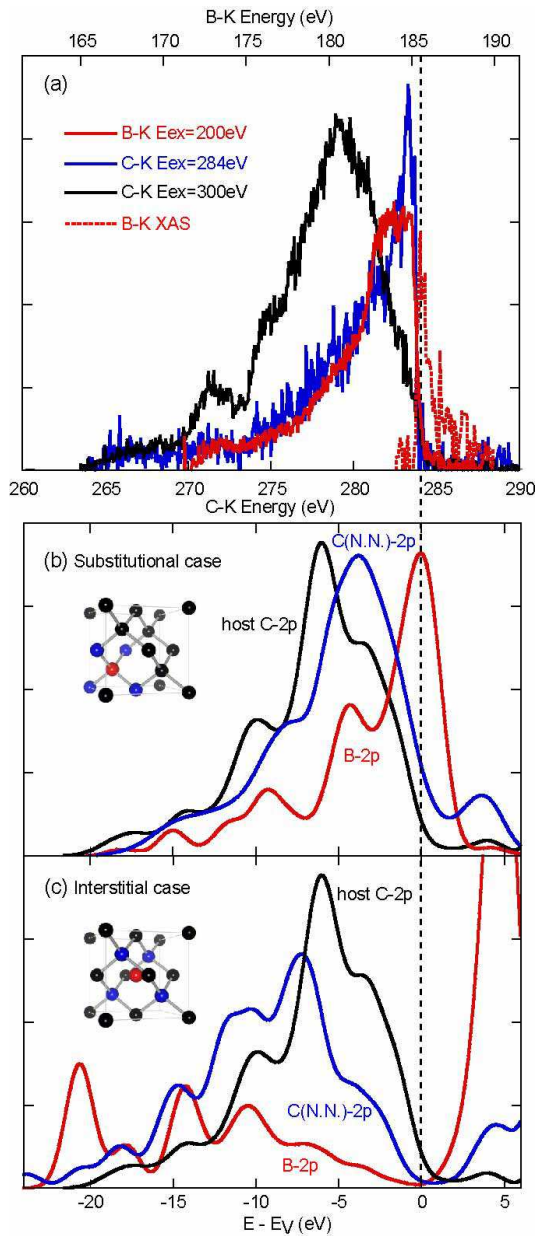}
\caption{\label{Fig5}
Comparison of observed B-K XAS and XES, and C-K XES spectra with DVX$\alpha$ simulations.  
(a) Observed XAS and XES spectra, (b) DVX$\alpha$-results of substitutional case, 
and (c) DVX$\alpha$-results of interstitial case.  In the inset pictures of (b) and (c), 
red and blue ball represent boron and N.N. carbon atoms, respectively.  }
\end{figure}

The results are shown in Fig.~\ref{Fig5}.  
Figure~\ref{Fig5}(a) shows experimental XES spectra of B-K with $E_{\rm ex}$ of 200 eV, 
C-K with $E_{\rm ex}$ of 284 eV and C-K with $E_{\rm ex}$ of 300 eV of B-diamond, 
corresponds to PDOS of B-2$p$, C(N.N.)-2$p$ and host C-2$p$, respectively.  
B-K XAS spectra of B-diamond near the Fermi level are also shown (red dashed line).  
The horizontal axes for B-K and C-K are shifted in order that the Fermi level is in 
agreement with each other.  
One can see that there is a good agreement in between B-2$p$ and C(N.N.)-2$p$ PDOS, 
which means strong hybridization between these orbitals.  
Figures~\ref{Fig5}(b) and (c) show the results in substitutional and interstitial cases.  
The origin of the energy in DVX$\alpha$ is set to the maximum energy of the electron in the 
occupied states, i.e., top of the V.B.  
The PDOS of C(N.N.)-2$p$ (blue line) is derived as averaged PDOS of four 
N.N. carbon atoms [the four blue balls in the inset in Figs.~\ref{Fig5}(b) and (c)].  
The PDOS of host C-2$p$ (black line) is the averaged one of a few carbon atoms near the 
center of the cluster but far from the dopant boron, which is in good agreement with 
the result of non-doped cluster,  C$_{184}$.  
All the calculated PDOS's are convoluted using gaussian function with the experimental 
width.  
In the substitutional case [Fig.~\ref{Fig5}(b)], PDOS of B-2$p$ (red line) shows a large peak 
around $E-E_{\rm V}$=0 eV, and PDOS of C(N.N.)-2$p$ possesses a main peak 
around $E-E_{\rm V}$=-3.8 eV.  
It is worth while to notice that the considerable amounts of state around 
$E_{\rm V}$ are appeared in both PDOS's of B-2$p$ and C(N.N.)-2$p$.  
It is consistent with the experimental results that the observed PDOS's of B-2$p$ and 
C(N.N.)-2$p$ also shows main states (peak) in higher energy side from the PDOS of host C-2$p$.  
Furthermore, the broad-beat structures of the B-2$p$ PDOS's tail in the low energy side 
agrees with those of C(N.N.)-2$p$ PDOS, which is consistent with the strong hybridization between 
dopant B-2$p$ and C(N.N.)-2$p$ observed experimentally [inset of Fig.~\ref{Fig4}(b) and Fig.~\ref{Fig5}(a)].   
The result in the substitutional case is consistent with the experimental one.  
On the other hand, in the interstitial case, both the PDOS's of B-2$p$ and C(N.N.)-2$p$ shift 
and broaden toward the low energy side.  
Especially large in-gap states for both B-2$p$ and C(N.N.)-2$p$ PDOS's are appeared with the threshold 
energy of about $E-E_{\rm V}\sim$+3 eV.  
There are almost no states near the $E_{\rm V}$, which does not support the experimental results.  
These results of DVX$\alpha$ suggest that dopant boron replace the carbon sites in these concentration 
region consistently with the previous report.\cite{werner}  
\section{Conclusions}
X-ray absorption (XAS) and emission (XES) spectra at the B-K and C-K edges have been 
performed on metallic ($\sim$1at\%B) and semiconducting ($\sim$0.1at\%B and N) doped-diamond films.  
Both B-K XAS and XES spectra show metallic partial density of states (PDOS) with the 
Fermi energy of 185.3 eV, and there is no apparent boron-concentration dependence 
in contrast to the different electric property.  
In C-K XAS spectrum of metallic B-diamond, the impurity state ascribed to boron is clearly 
observed near the Fermi level.  
The Fermi energy is found to be almost same with the top of the valence band of non-doped 
diamond, $E_{\rm V}$, 283.9 eV.  
The C-K XAS of semiconducting BN-diamond shows both the B-induced shallow level and 
N-induced deep-and-broad levels as the in-gap states, in which the shallow level is 
in good agreement with the activation energy ($E_{\rm a}$=0.37 eV) estimated from the 
temperature dependence of the conductivity, 
namely the change in C-2$p$ PDOS of impurity-induced metallization is directly observed.  
The electronic property of these diamonds is mainly attributed to the electronic structure of 
C-2$p$ near the Fermi level.  
The observed XAS and XES spectra are compared with the DVX$\alpha$ cluster calculations.  
The DVX$\alpha$ result supports the strong  hybridization between B-2$p$ and C-2$p$ observed in 
XAS and XES spectra, and suggests that borons in diamond occupy the substitutional site in the present 
doping range between 0.1at\%B and 1at\%B in diamond, rather than interstitial site. 
\section*{Acknowledgment}
We express our thanks to Dr. Y. Muramatsu of Japan Atomic Energy Research Institute 
(JAERI), Prof. K. Kuroki of University of Electro-Communications and Prof. T. Oguchi 
of Hiroshima University for useful discussions.  
J.N. wishes to acknowledge to Dr. J.D. Denlinger of the Advanced Light Source (ALS)
at Lawrence Berkeley National Laboratory (LBNL) for the experimental support.  
This work was performed under the approval of ALS-LBNL, proposal No. ALS-00931.  
ALS is supported by the Director, Office of Science, Office of Basic Energy Sciences, 
Materials Sciences Division, of the U.S. Department of Energy 
under Contract No. DE-AC03-76SF00098 at LBNL.
%


\begin{thebibliography}{99}
\bibitem{slit} The intensity of the peak at 284.2 eV in XAS spectrum is weak, 
to increase the flux, the slits of the monochromator were increased so that the resolution 
is same as the FWHM of the 284.2 eV peak ($\sim$0.6 eV).
\end{thebibliography}

\end{document}